\documentclass[%
 reprint,
 amsmath,amssymb,
 aps,
floatfix,showpacs
]{revtex4-2}
\usepackage{graphicx}
\usepackage{dcolumn}
\usepackage{bm}
\usepackage{amsmath}

\usepackage[colorlinks,linkcolor=blue,citecolor=blue,urlcolor=blue,hyperindex,breaklinks]{hyperref}
\usepackage{subfigure}

\begin{document}

\title{Tight and attainable quantum speed limit for open systems}
\author{Zi-yi Mai}
\author{Chang-shui Yu}
\email{ycs@dlut.edu.cn}
\date{\today}
\affiliation{School of Physics, Dalian University of Technology, Dalian
116024, P.R. China}
\begin{abstract}
We develop an intuitive geometric picture of quantum states, define a particular state distance, and derive a quantum speed limit (QSL) for open systems. Our QSL is attainable
because any initial state can be driven to a final state by the particular dynamics along the geodesic.  We present the general condition for dynamics along the geodesic for our QSL. As evidence, we consider the generalized amplitude damping dynamics and
the dephasing dynamics to demonstrate the attainability. In addition, we
also compare our QSL with others by strict analytic processes as well as numerical illustrations, and show our QSL is tight in many cases.  It indicates that our work is significant in tightening the bound of evolution time.

\end{abstract}
\maketitle

\section{Introduction}
Quantum speed limit (QSL) (or equivalently to call the quantum speed limit time (QSLT)) is an important feature of
a dynamical system, which mainly characterizes the minimal time required for a
state evolving to a target state. It is a constrained optimization problem important in quantum metrology \cite{1,2,3}, quantum optimal control \cite{4,5, PhysRevA.84.022305, Poggi_2013},  quantum information processing \cite{6, PhysRevResearch.2.013161}. Recently, it's considered a meaningful index for a given quantum system to evaluate its dynamics characteristics involving robustness \cite{PhysRevA.102.042606}, non-Markovianity \cite{PhysRevA.91.032112,PhysRevA.93.020105}, upper bound of changing rate of expected value of observable \cite{PhysRevX.12.011038}, decoherence time \cite{PhysRevX.6.021031, PhysRevA.93.052331, PhysRevLett.118.140403, PhysRevLett.119.130401},  interaction speed in spin system \cite{PhysRevLett.102.017204, PhysRevA.70.022308} and changing rate of phase \cite{PhysRevLett.127.100404} and so on \cite{osti_4459732, PhysRevE.103.032105}. Besides, the quantum speed limit is widely used to explore the intrinsic nature of physical systems, such as for the many-body system \cite{PhysRevX.9.011034}, ultracold atomic system \cite{PhysRevLett.126.180603}, non-Hermitian system \cite{PhysRevA.86.064104} and entanglement \cite{PhysRevA.91.022102, PhysRevA.78.042305, PhysRevA.89.012307, Zander_2007,Wu_2015, PhysRevX.9.011034}. For the application fields, studies of the quantum speed limit are involved in machine learning \cite{PhysRevA.97.052333}, quantum measurement \cite{Garcia_Pintos_2019} and thermometry \cite{Campbell_2018}.

QSL was first addressed for a unitary evolution from a pure state to its orthogonal state by Mandelstam and Tamm \cite{7}, who presented the famous
time-Energy uncertainty (MT bound) $\tau _{MT}^{\bot }=\pi /(2\Delta E)$, where $(\Delta E)^{2}=\langle H^{2}\rangle -\langle H\rangle ^{2}$ stands
for the variance of Hamiltonian of the system \cite{Deffner_2013}. Later, Margolus and Levitin \cite{8} established another bound (ML bound) of the
unitary evolution between pure orthogonal states as $\tau _{ML}^{\bot }=\pi/2E$ based on the average energy $E$ \cite{Deffner_2013}. A tighter bound was obtained by the
combination of MT and ML bounds as $\tau _{MT-ML}^{\bot }=\pi /(2\min\{E,\Delta E\})$ \cite{Vittorio_Giovannetti_2004}. Giovannetti et al. generalized MT and
ML bounds to the mixed initial state \cite{PhysRevA.67.052109}. However, a deeper understanding of the QSL could count on the geometrical perspective first developed by Anandan and Aharanov for the MT bound with time-dependent
Hamiltonian in terms of the Fubini-Study metric on the pure-state space \cite{PhysRevLett.65.1697}. Up to now, various geometrical distances have been exploited to develop QSL for
density matrices \cite{russell2014geometrical,10,11,12,13,14,15,zhang2014quantum,17, Deffner_2013,19,20,21,sun2015quantum, PhysRevA.93.052331, PhysRevX.12.011038, PhysRevLett.123.180403, PhysRevA.86.016101}. Considering the inevitable contact with environments, the QSL has also been developed for open system \cite{PhysRevA.78.012308} based on different metrics such as quantum Fisher
information \cite{25}, relative purity \cite{21}, and the MT bound
and the ML bound have been extended to open systems in terms of a
geometric way \cite{deffner2013quantum}. In addition, QSL is also characterized
based on quantum resource theory \cite{campaioli2022resource}. It is even shown that the speed limit is not a unique phenomenon in quantum systems \cite{PhysRevLett.120.070402, PhysRevLett.120.070401}. Every QSL could have its significance in that it gives a potentially different understanding of the bound on the evolution time of a system. The most typical examples are the MT and ML QSLs which bound the evolution time by the fluctuation of energy and the average energy, respectively. In this sense, it is important to establish a distinguished QSL.

The tightness and the attainability are key aspects of a good QSL bound, which strongly depends on the different understanding perspectives of QSL \cite{frey2016quantum,deffner2017quantum}. If
the dynamics (Hamiltonian or Lindblad) is fixed, the QSLT bounds the minimum
evolution time between a pair of states with a given `distance.' If the state
`distance' is given, the tight QSL bound means that dynamics drive the given initial state to the final state with the minimum time. MT and ML bounds are attainable for a unitary evolution if the initial state is
the equal-weight superposition of two eigenstates of the Hamiltonian with
zero ground-state energy \cite{8}. Ref. \cite{15} generalized the tight bound for the unitary case to the mixed states, Ref. \cite{shao2020operational} verifies this bound is attainable for any dimension.  Ref. \cite{campaioli2019tight} proposed a QSLT bound attainable for dephasing and depolarized channels. For a tight bound, many papers focus on combining different QSLs.

In this paper, we establish a tight and attainable QSLT for open systems in terms of a
geometric approach. Similar to the Bloch representation of quantum states, we develop an intuitive geometric picture of quantum states. All the states are mapped to the surface of the high-dimensional sphere.
In this picture, we derive a QSL for an open quantum system by a particularly-defined state distance. Our QSL is attainable in that for any given initial state, one can always find a dynamics to drive the initial
state to the final state along the geodesic. In particular, we present a general condition for dynamics along the geodesic. The generalized amplitude damping dynamics and the dephasing dynamics evidence the attainability. In addition, we compare our QSL and the one  in Ref. \cite{campaioli2019tight} by considering the unitary evolution of pure states and the particular amplitude damping dynamics. It is shown that our QSL is tight.  In addition, numerical examples show that the QSLT of Ref. \cite{campaioli2019tight} is tighter than ours in many cases,  which implies the combination of the two QSLs is necessary.
The paper is organized as follows.  We first propose the intuitive geometric picture of quantum states and present our QSL. Then we arrive at the general condition for dynamics along the geodesic, and then we give concrete examples to demonstrate the attainability of our QSL. Finally, we show the tightness of our QSL by comparing particular dynamics.

\section{Quantum speed limit}

For an open system, the evolution of the
quantum state $\rho _{t}$ is governed by the general master equation as%
\begin{equation}
\dot{\rho}_{t}=\mathcal{L}_{t}\left( \rho _{t}\right) ,  \label{1}
\end{equation}%
where $\mathcal{L}_{t}\left( \cdot \right) $ denotes a general dissipator of
the system and the subscript $t$ indicates the potential dependence on time,
in particular, we don't specify whether $\mathcal{L}_{t}\left( \cdot \right)
$ is Lindblad or not. Let $P_{t}=P\left( \rho _{t}\right) =\rho _{t}/%
\sqrt{Tr\rho _{t}^{2}}$, then for any pair of $P_{t}$ and $P_{t}^{\prime }$
we can define \ \
\begin{equation}
D(P_{t}||P_{t}^{\prime })=\arccos \langle P_{t},P_{t}^{\prime }\rangle
\end{equation}%
based on the Hilbert-Schmidt distance $\langle P_{t},P_{t}^{\prime }\rangle
=TrP_{t}^{\dag }P_{t}^{\prime }$.  Based on \textit{Schoeberg's Theorem}
 \cite{schoenberg1938metric}, which is firstly introduced by Ref. \cite{PhysRevA.78.052330} to tackle with distance function equipped by the metric space consisted of density matrices, one can easily prove that $D(P_{t}||P_{t}^{%
\prime })$ is a good distance. Thus all $P_{t}$ can form a metric space $%
S\left( \mathcal{H}\right) $ with respect to the distance $%
D(P_{t}||P_{t}^{\prime })$. It is obvious that $D(P_{t}||P_{t}^{\prime })$
for a pair of density matrices $\rho $ and $\sigma $ can be explicitly
written as
\begin{equation}
D(\rho ||\sigma )=\arccos \mathcal{F}_{GM}(\rho ,\sigma ),  \label{DGM}
\end{equation}%
where $\mathcal{F}_{GM}(\rho ,\sigma )=Tr\rho \sigma /(\sqrt{Tr\rho ^{2}}%
\sqrt{Tr\sigma ^{2}})$ is the alternative fidelity introduced in Ref. \cite%
{27}. The alternative fidelity is also used in a different way for QSLT in Ref. \cite{PRXQuantum.2.040349}.

To get the QSLT, we need the differential form of the distance $D(\rho
||\sigma )$. Considering the infinitesimal evolution $\rho _{t}\longmapsto
\rho _{t}+d\rho _{t}$, the distance reads
\begin{equation}
ds=D(\rho _{t}||\rho _{t}+d\rho _{t})=\arccos \frac{Tr\rho _{t}(\rho
_{t}+d\rho _{t})}{\sqrt{Tr\rho _{t}^{2}}\sqrt{Tr(\rho _{t}+d\rho _{t})^{2}}}.
\end{equation}%
A direct deformation gives $\frac{Tr\rho _{t}(\rho _{t}+d\rho _{t})}{\sqrt{%
Tr\rho _{t}^{2}}\sqrt{Tr(\rho _{t}+d\rho _{t})^{2}}}=\cos ds=1-\frac{ds^{2}}{%
2}$, which indicates
\begin{equation}\label{ds}
ds^{2}=2(1-\frac{Tr\rho _{t}(\rho _{t}+d\rho _{t})}{\sqrt{Tr\rho _{t}^{2}}%
\sqrt{Tr(\rho _{t}+d\rho _{t})^{2}}}).
\end{equation}%
Under the condition $d\rho _{t}\longmapsto 0$, we can expand $\frac{1}{\sqrt{%
Tr(\rho _{t}+d\rho _{t})^{2}}}$ to the second order:
\begin{equation}\label{expand}
  \frac{1}{\sqrt{%
Tr(\rho _{t}+d\rho _{t})^{2}}}=\frac{1-\frac{Trd\rho_t^2}{2Tr\rho_t^2}-\frac{Tr\rho_td\rho_t}{Tr(\rho_t)^2}+\frac{3(Tr\rho_td\rho_t)^2}{2(Tr\rho_t)^2}}{\sqrt{Tr\rho_t^2}}
\end{equation}

Substituting Eq. (\ref{expand}) into Eq. (\ref{ds}), then we can immediately
obtain the metric as
\begin{equation}\label{metric_1}
ds^{2}=\frac{Tr(d\rho _{t})^{2}Tr\rho _{t}^{2}-(Tr\rho _{t}d\rho _{t})^{2}}{%
(Tr\rho _{t}^{2})^{2}}.
\end{equation}%
Denote $\mathcal{L}_{t}(P_{t})=\dot{\rho}_{t}/\sqrt{Tr\rho _{t}^{2}}$ if not
confused(especially for the Lindbladian \cite{10.1063/1.5115323}, a form of $\mathcal{L}_t(P_t)$ is reasonable because $1/\sqrt{Tr\rho_t^2}$ is just a real number which is commutative with any operator), then the metric given in Eq. (\ref{metric_1}) turns to form of the
Fubini-Study metric as
\begin{equation}
(ds/dt)^{2}=\langle \mathcal{L}_{t}(P_{t}),\mathcal{L}_{t}(P_{t})\rangle
-\langle P_{t},\mathcal{L}_{t}(P_{t})\rangle ^{2}.  \label{metric}
\end{equation}%
For infinitesimal $dt$, we have
\begin{eqnarray}
P_{t+dt} &=&\frac{\rho _{t}+d\rho _{t}}{\sqrt{Tr(\rho _{t}+d\rho _{t})^{2}}}
\notag \\
&=&P_{t}+\mathcal{L}_{t}(P_{t})dt-P_{t}\langle P_{t},\mathcal{L}%
_{t}(P_{t})\rangle dt.
\end{eqnarray}%
According to $\dot{P}_{t}=\frac{P_{t+dt}-P_{t}}{dt}$, one can arrive at
\begin{equation}
\langle \dot{P}_{t},\dot{P}_{t}\rangle =\langle \mathcal{L}_{t}(P_{t}),%
\mathcal{L}_{t}(P_{t})\rangle -\langle P_{t},\mathcal{L}_{t}(P_{t})\rangle
^{2}.  \label{inner}
\end{equation}%
Comparing Eq. (\ref{metric}) and Eq. (\ref{inner}), we have
\begin{equation}
v_{t}^{2}=(ds/dt)^{2}=\langle \dot{P}_{t},\dot{P}_{t}\rangle .
\end{equation}

In the above metric space, we can derive a speed limit based on the metric
Eq. (\ref{inner}) and the distance Eq. (\ref{DGM}) are as follows.

\textit{Theorem 1}.-The minimal time for a given state $\rho _{0}$ to evolve
to the state $\rho _{\tau }$ subject to the dynamics Eq. (\ref{1}) is lower
bounded by
\begin{equation}
\tau _{qsl}=\frac{\arccos \langle P_{0},P_{\tau }\rangle }{\frac{1}{\tau }%
\int\nolimits_{0}^{\tau }\sqrt{\langle \dot{P}_{t},\dot{P}_{t}\rangle }dt}
\label{QSLT1}
\end{equation}%
with $P_{0}=\rho _{0}/\sqrt{Tr\rho _{0}^{2}}$ and $P_{\tau }=\rho _{\tau }/%
\sqrt{Tr\rho _{\tau }^{2}}$.

\textit{Proof}.-Based on the distance, one can find that
\begin{gather}
\arccos \langle P_{0},P_{\tau }\rangle =D(\rho _{0}||\rho _{\tau })  \notag
\\
\leq \sum_{t=0}^{\tau }D(\rho _{t}||\rho _{t}+d\rho _{t})=\int_{0}^{\tau
}\left\vert \frac{ds}{dt}\right\vert dt  \notag \\
=\int\nolimits_{0}^{\tau }\sqrt{\langle \dot{P}_{t},\dot{P}_{t}\rangle }dt,
\label{pro11}
\end{gather}%
which directly leads to $\tau \geq \tau _{qsl}$ as given in Eq. (\ref{QSLT1}%
).\hfill{} $\square $

Now we'd like to give an intuitive understanding of the map between this metric space and
the Bloch representation.  As is given in Fig. 1. the states in the metric space
form a spherical crown and are one-to-one mapped to the bottom surface of the hemispherical surface, which is geometrically the same as the circular section across the center of the
Bloch sphere and the two points $\rho $ and $\sigma $. The apex of the
the spherical crown is the maximally mixed state. The latitude of the bottom
the surface of the spherical crown is determined by the intersection angle of a
pure state and the maximally mixed state or the dimension of the state
space. All the states of the same mixedness are distributed on the same
latitude, which especially implies unitary evolution along the latitude. The
evolution of purely reducing mixedness will undergo longitude. It can be noticed that for the evolution trajectories tracing geodesics in the Bloch sphere which is equipped by the Euclidean distance, it also traces a geodesics in our metric space.\begin{figure}[tbp]
\includegraphics[width=8cm]{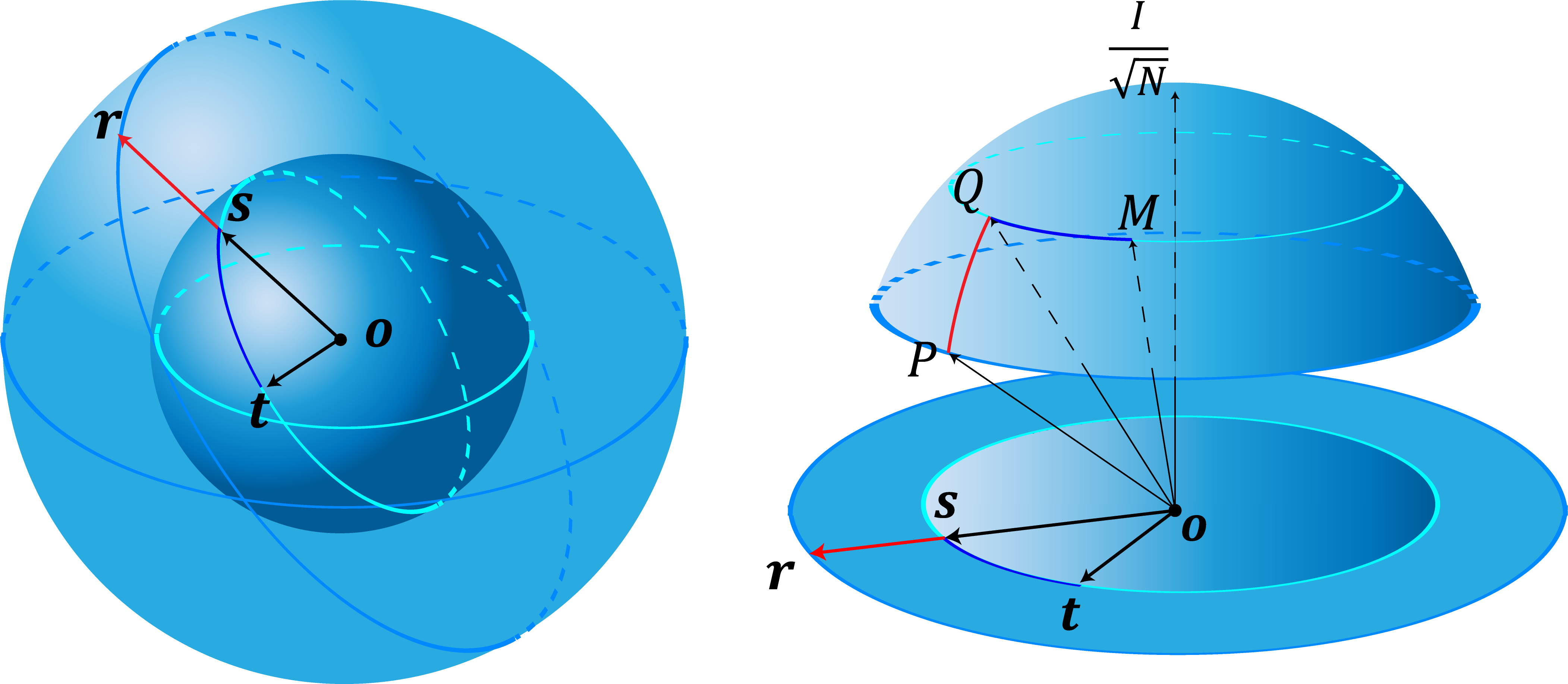}
\caption{The geometric picture of quantum states. The left figure is the Bloch sphere, and the right is our new geometric picture. The arrow $\bm{r}$ in the Bloch sphere is the Bloch vector, representing the corresponding density matrix $\rho$ according to  $\rho=\frac{1}{N}\left(I+\sqrt{\frac{N(N-1)}{2}} \bm{r}\cdot\bm{A}\right)$, $\bm{s}$ and $\bm{t}$ denotes the correponding Bloch vector of the state $\sigma$ and $\epsilon$, respectively, where $\bm{A}=\left(A_1, ..., A_N\right)$ is a Lie algebra for SU($N$) \cite{15}.
Focusing the cross-section, which involves the Bloch vectors $\bm{r}$, $\bm{s}$, $\bm{t}$ and the original point $\bm{o}$ (the maximally mixed state), the cross-section can be reshaped as a crown (in the right picture) of the unit sphere in the matrix space which equips with the Hilbert-Schmidt inner product. Equivalently, the crown consists of the normalized density matrices: $P=\rho/\sqrt{Tr\rho^2}$, similarly, $Q$ and $M$ are defined for $\sigma$ and $\epsilon$, respectively. Overall, the new metric space is actually an extension of the one-dimensional cross-section of the Bloch sphere to the three-dimensional space.
Therefore, the unitary trajectories tracing the great circle on the surface of the Bloch sphere,  correspond to the latitude on the spherical crown in the new metric space (the blue trajectories).  The depolarization trajectories that evolve along the radial direction of the Bloch sphere correspond to the longitude in the new metric space  (the red trajectories).
}
\end{figure}

\section{Attainability and Tightness}
\textit{Attainability.-}It is easy to find that the quantum speed limit time
$\tau _{qsl}$ is expressed by the distance $\arccos \langle P_{0},P_{\tau
}\rangle $ divided by the average evolution speed $\left\vert \bar{v}%
_{t}\right\vert =\frac{1}{\tau }\int\nolimits_{0}^{\tau }\sqrt{\langle \dot{P%
}_{t},\dot{P}_{t}\rangle }dt$. Next, we will show that the QSL time
presented in Theorem 1 can be attainable. Namely, given a distance and the
average speed, one can always find a pair of quantum states and a
corresponding dynamics such that the practical evolution time is exactly the
QSL time.

\textit{Theorem 2}.-The evolution of $\rho _{t}$ from a given initial state $%
\rho _{0}$ to a final state $\rho _{\tau }$ along the geodesic can be
written as%
\begin{equation}
\dot{\rho}_{t}=\frac{\dot{\beta}(t)}{\beta (t)}\left( \rho _{t}-\rho
_{0}\right) ,  \label{geoeqn}
\end{equation}%
and hence the geodesic is
\begin{equation}
\rho _{t}=(1-\beta (t))\rho _{0}+\beta (t)\rho _{\tau },  \label{geodesic}
\end{equation}%
where $\beta (t)$ is a monotonic function with $\beta (0)=0$ and $\beta
(\tau )=1$.

The proof is given in Appendix A.  Ref. \cite{PhysRevA.25.3206} shows that the form of (\ref{geoeqn}) can describe the behavior of atomic decay.  It can be verified Eq. (\ref{geodesic}) is also the geodesics of the bound from Ref. \cite{campaioli2019tight}. In fact, one can easily verified that, the arbitrariness of $\beta_t$ and $\rho_\tau$ guarantee a highly freedom of form of the geodesics:
\begin{equation}\label{geodesic_1}
  \rho_t=\rho_0+\beta(t)C,
\end{equation}
where $C$ is arbitrary traceless hermitian matrix.

Theorem 2 explicitly indicates the general form of the geodesic. In
particular, $\rho _{\tau }=\frac{1}{N}$ corresponds to the longitude
equation. Eq. (\ref{geoeqn}) means the density matrix that evolves along a
geodesic or the QSLT is attainable. However, it can be shown that bound (\ref{QSLT1}) is impossible to be saturated for any unitary case by making a comparison to the bound $\tau_\Phi$ from Ref. \cite{15}:
\begin{equation}
  \tau\geq \tau_\Phi=\frac{\sqrt{2}arccos\sqrt{\langle P_0,P_\tau\rangle}}{\frac{1}{\tau}\int_0^\tau\sqrt{\langle \dot{P}_t,\dot{P}_t\rangle }dt}>\tau_{qsl},\ \tau\neq 0\label{unitary_case}
\end{equation}
(\ref{unitary_case}) derived from the monotone decreasing function $f(x)=\sqrt{2}arccosx-arccosx^2>f(1)=0$ when $0<x<1$.

\textit{Examples.-}Considering an $N$-level system coupling to a heat bath with $b_k$ denoting the annihilator of its $k$th mode,    the Hamiltonian for the total system is $H=\sum_{i=0}^NE_i\left\vert i\right\rangle\left\langle i\right\vert +\sum_k\omega_kb_k^\dagger b_k+\sum_{ik}(g_k\sigma_+^ib_k+h.c)$, where $E_i$ is the energy of $i$th energy level,  and  $\sigma
_{-}^{i}=\left\vert 0\right\rangle \left\langle i\right\vert $ and $\sigma
_{+}^{i}=\left\vert i\right\rangle \left\langle 0\right\vert $ denote the transition operators.  Following Ref. \cite{breuer2002theory},  one can obtain the dynamics for the reduced system as
\begin{equation}
\dot{\rho _{t}}=-\sum_ki[\frac{s^k_t}{2}\left\vert k\right\rangle\left\langle k\right\vert, \rho_t]+\frac{\gamma ^k_{t}}{2}(2\sigma _{-}^{k}\rho _{t}\sigma
_{+}^{k}-\sigma _{+}^{k}\sigma _{-}^{k}\rho _{t}-\rho _{t}\sigma
_{+}^{k}\sigma _{-}^{k}),  \label{mas1}
\end{equation}%
where $s^k_t$ is the time-dependent Lamb shift and $\gamma^k_t$ represents the time-dependent decay rate.  This equation describes the generalized amplitude dampling dynamics. Let the initial state be
 $\rho
=\sum\limits_{i=0}\lambda _{i}\left\vert i\right\rangle \left\langle
i\right\vert $ and suppose $\gamma^k_t\equiv \gamma_t$,
the density matrix $\rho _{t}$ can be solved as
\begin{equation}
\rho _{t}=\left( 1-\sum\limits_{i\neq 0}\lambda _{i}q_{t}\right) \left\vert
0\right\rangle \left\langle 0\right\vert +\sum\limits_{i}\lambda
_{i}q_{t}\left\vert i\right\rangle \left\langle i\right\vert  \label{statet}
\end{equation}%
with $q_{t}=e^{-\int\nolimits_{0}^{t}\gamma _{t^{\prime }}dt^{\prime }}$.
Derivation of $\rho _{t}$ in Eq. (\ref{statet}), we have $\dot{\rho _{t}}=%
\dot{q}_{t}\left( \sum\limits_{i}\lambda _{i}\left\vert i\right\rangle
\left\langle i\right\vert \right) $, $\lambda _{0}=1-\sum\limits_{i\neq
0}\lambda _{i}$, which means the QSLT will be attainable due to theorem 2.
To explicitly show it, let's substitute Eq. (\ref{mas1}) and Eq. (%
\ref{statet}) into Eq. (\ref{QSLT1}), one can immediately find that in the
duration $\tau $, the distance in terms of the average evolution speed is
\begin{eqnarray}\label{jc02}
\tau \bar{v}_{t} &=&\int\nolimits_{0}^{\tau }\frac{|\frac{dq_{t}}{dt}|c}{%
1+aq_{t}^{2}-2bq_{t}}dt  \notag \\
&=&\left\vert\arctan \frac{a|q_{\tau }|-b}{c}-\arctan \frac{a-b}{c}\right\vert,
\end{eqnarray}%
where $c=\sqrt{\sum\limits_{i}\lambda _{i}^{2}}$, $b=\sum_{i\neq 0}\lambda
_{i}$,  $a=b^{2}+c^{2}$ and we suppose $q_t$ is monotonic. The distance away from the initial state $\rho
_{0}$ is%
\begin{equation}
D(\rho _{\tau }||\rho _{0})=\arccos \frac{1-b(|q_{\tau }|\text{+}%
1)+a|q_{\tau }|}{f(q_{0})f(q_{\tau })}
\end{equation}%
with $f(q_{\tau })=\sqrt{1-2b|q_{\tau }|^{2}+a|q_{\tau }|^{2}}$ and $q_{0}=1$%
. It is easy to find that $D(\rho _{\tau }||\rho _{0})=\tau \bar{v}_{t}$,
which directly shows the quantum speed limit time is consistent with the
practical evolution time, $\tau _{qsl}=\frac{D(\rho _{\tau }||\rho _{0})}{%
\bar{v}_{t}}=\tau $.

The other attainable case is the dephasing dynamics. Suppose the above $N$-level system  undergoes an environment consisting of multiple reservoirs with each two energy levels driven by an individual reservoir.  Let the $j$th and the $k$th levels interact with the reservoir as $H_{jk}=\sum_{\nu }\sigma^z_{jk}(g_\nu b_\nu^\dagger+g_\nu^*b_\nu)$,  where $\sigma^z_{jk}=\left\vert j\right\rangle\left\langle j\right\vert-\left\vert k\right\rangle\left\langle k\right\vert$ for $j>k$,  $g_\nu$ is the coupling strength, and $b_\nu$ is any operator of the reservoir corresponding to the  $j$th and  $k$th levels.  Consider the time evolution of the system \cite{breuer2002theory},  for any initial state $\rho(0)$ one will get the final state as
\begin{equation}
\rho(t)=\sum_{mn}\rho_{mn}(0)\left\vert m\right\rangle\left\langle n\right\vert Tr_B\left\{V_n^{-1}(t)V_m(t)\rho^{mn}_{B}(0)\right\},
\end{equation}
where $V_m(t)$ is derived from the time-evolution operator performing on the state $\left\vert m\right\rangle$,  and $\rho^{mn}_B(0)$ is the potential initial state of the reservoir corresponding to the  $m$th and  $n$th levels.
Define the decay rates as
\begin{equation}\label{mas_3}
 \Gamma_{mn}(t)=\ln{Tr_B\left\{V_n^{-1}(t)V_m(t)\rho^{mn}_{B}(0)\right\}}\equiv -\gamma_t
\end{equation}
independent of $mn$, then the final state can be written as
\begin{equation}
\rho _{t}=\sum_{i}\rho _{ii}(0)\left\vert i\right\rangle \left\langle
i\right\vert +e^{-\gamma _{t}}\sum_{j\neq k}\rho _{jk}(0)\left\vert
j\right\rangle \left\langle k\right\vert. \label{deph}
\end{equation}%
The derivative of $\rho _{t}$ reads
\begin{equation}
\dot{\rho _{t}}=-\frac{d\gamma _{t}}{dt}e^{-\gamma _{t}}\sum_{j\neq k}\rho
_{jk}(0)\left\vert j\right\rangle \left\langle k\right\vert .  \label{depd}
\end{equation}%
It is evident that Eq. (\ref{depd}) has the same form as that in theorem 2,
so the QSLT is attainable.

Again, let's substitute Eq. (\ref{deph}) and Eq. (\ref{depd}) into Eq. (\ref%
{QSLT1}), we can express the distance based on the average evolution speed
as
\begin{eqnarray}\label{dephasing02}
\tau \bar{v}_{t} &=&\int\nolimits_{0}^{\tau }\left\vert \frac{d\gamma _{t}}{%
dt}\right\vert \frac{Re^{-\gamma _{t}}}{1+R^{2}e^{-2\gamma _{t}}}dt  \notag
\\
&=&\arctan R-\arctan \left( e^{-\gamma _{\tau }}R\right)
\end{eqnarray}%
where $R=\sqrt{\frac{\sum_{j\neq k}|\rho _{jk}|^{2}}{\sum_{i}\rho _{ii}^{2}}}$ and $\gamma_t$ is supposed to be monotonic. 
The distance away from the initial state is%
\begin{equation}
D(\rho ||\sigma )=\arccos \frac{F_{\tau }^{2}\left( 2,1\right) }{F_{\tau
}(2,0)F_{\tau }(2,2)}
\end{equation}%
with $F_{\tau }\left( k,s\right) =\sqrt{1+R^{k}e^{-s\gamma _{\tau }}}$. A
further simplification can show that $D(\rho ||\sigma )=\tau \bar{v}_{t}$,
which means the QSL time $\tau _{qsl}=\frac{D(\rho ||\sigma )}{\bar{v}_{t}}%
=\tau $.

At first, we would like to emphasize that in the two examples  we choose the particular dynamics and the initial states to demonstrate the attainability.  In fact, both the Hamiltonian and the initial states can change the attainability.  In Appendix B,   we present an example of qubit system to demonstrate the deviation of the evolution trajectory from the geodesic due to different Hamiltonian and initial states.

 In addition, we don't specify the explicit form of the decay rates except for monotonicity, the non-monotonicity or the divergence of $\gamma_t$ will force the evolution trajectory oscillates back and forth over the geodesics,  and lead to $\tau_{qsl}<\tau$, namely,  the evolution trajectory deviates from the geodesics.  For example,  if the decay rate takes
\begin{equation}
\gamma _{t}=\frac{2\gamma _{0}\lambda \sinh (\delta t/2)}{\delta \cosh
(\delta t/2)+\lambda \sinh (\delta t/2)},
\end{equation}
where  $\delta =\sqrt{\lambda ^{2}-2\gamma _{0}\lambda }$, and $\lambda $ and $\gamma _{0}$ represent the spectral width and coupling strength, respectively. If
the parameter $\delta  $ is real,
i.e., $\gamma _{0}\leq \lambda /2$, the dynamics is Markovian, which implies
a relatively weak coupling, and the decay rate can be taken as a constant $%
\gamma _{t}=\gamma _{0}$ for $\gamma _{0}\ll \lambda $. Conversely, if $%
\gamma _{0}\geq \lambda /2$, it means the stronger coupling described by the
non-Markovian dynamics, which  leads to non-monotonic $\gamma_t$. The evolution trajectory is not the geodesics,  which is similar to the unsaturated effect of QSL bounds for the non-Markovian dynamics reported by the previous works \cite{deffner2013quantum,PhysRevA.93.020105,PhysRevA.96.012105}.

The above two dynamics indicate the attainability of our QSL time in any
dimensional state space. It is obvious that the farthest evolution is
governed by the nonunitary dynamics instead of the unitary process. If we
restrict the system to undergo the unitary evolution subject to the
Hamiltonian $H_{t}$, the average speed $\bar{v}_{t}$ will be reduced to $%
\bar{v}_{t}=\frac{1}{\tau }\int\nolimits_{0}^{\tau }\sqrt{-\frac{1}{\hbar
^{2}}Tr\{[P_{t},H_{t}]\}^{2}}dt$, which is the same as the speed $Q_\Phi$ in Ref. \cite%
{15}. However, one can find that the effective distance $D(\rho ||\sigma )$
is strictly larger than the distance $\Phi(\rho||\sigma)$ in Ref. \cite{15} in nontrivial dynamics, so the practical evolution is strictly larger than our presented QSL time $\tau
_{qsl}$. So our QSL cannot be attainable for a nontrivial unitary process.

\textit{Tightness.-} Tightness is an important question in the QSL, which depends on not only  the particular QSLT itself, but also the understanding of QSLT. For example, the MT and ML bounds for the unitary evolution can be the tightest, since they are obviously attainable for any pair of states as mentioned previously.  Of course, if we understand the QSLT in the sense that for any given initial state whether one can find a proper dynamics to drive the state evolve along the geodesic, then our obtained QSLT is also  the tightest since we have explicitly demonstrated the attainability.  However, in the general sense,   it is  quite hard to evalute the tightness of a QSTL for open systems,  because it is impossible to  exhausted all the potential evolution trajectories to demonstrate the tightness of a single QSLT or to compare the inifinite QSLTs due to their dependence on the evolution trajectory.  Therefore,  we follows the usual and feasible comparison approach in the literature \cite{pires2016generalized,deffner2013quantum,21,22,campaioli2019tight} namely, for given initial and final state, we compare two QSLTs subject to the same evolution trajectory.

Recently Ref.  \cite{campaioli2019tight} has presented a tight QSLT  \begin{equation}\label{QSL_modi}
  \tau_E=\frac{E(\rho_0||\rho_\tau)}{\frac{1}{\tau}\int_0^\tau dt \sqrt{Tr\dot{\rho}_t^2}}
\end{equation} with good tightness based on the Euclidean distance $E(\rho_0||\rho_\tau)=\sqrt{Tr(\rho_0-\rho_\tau)^2}$.  It has been shown that $\tau_E$ shares the same geodesic as our QSLT $\tau_{qsl}$.  Here we would like to compare our QSLT with $\tau_E$. 

Since  $TrA^2TrB^2\geq (TrAB)^2$ for any Hermitian operators $A$ and $B$ \cite{Jin}, one can easily find
\begin{equation}\label{speedlimit_ineq}
     \sqrt{Tr\dot{\rho}_t^2}  \leq\frac{\sqrt{Tr\dot{\rho}_t^2Tr\rho_t^2-\left(Tr\rho_t\dot{\rho}_t\right)^2} }{Tr\rho_t^2}.
\end{equation}
It is obvious that the left-hand side of the inequality (\ref{speedlimit_ineq}) is the evolution speed of the bound $\tau_E$, and the right-hand side is the evolution speed of our bound.  Because these two bounds are saturated by the same dynamics \cite{10},   integrating Eq. (\ref{speedlimit_ineq}) one will immediately arrive at
\begin{equation}\label{distance_ineq}
  \sqrt{Tr(\rho_0-\rho_\tau)^2}< \arccos\frac{Tr\rho_0\rho_\tau}{\sqrt{Tr\rho_0^2}\sqrt{Tr\rho_\tau^2}}.
\end{equation}
When we restrict the unitary evolution and the pure initial state,  the two sides of Eq. (\ref{speedlimit_ineq}) are identical, but inequality in Eq. (\ref{distance_ineq}) still holds,  which implies our bound shows $\tau_{qsl}>\tau_E$. In particular,   the continuity of QSLT promises that for some evolution trajectories closed to the unitary path for the pure state, our bound still shows preferable tightness compared with $\tau_E$.

\begin{figure}[htbp]
\includegraphics[width=6.0cm]{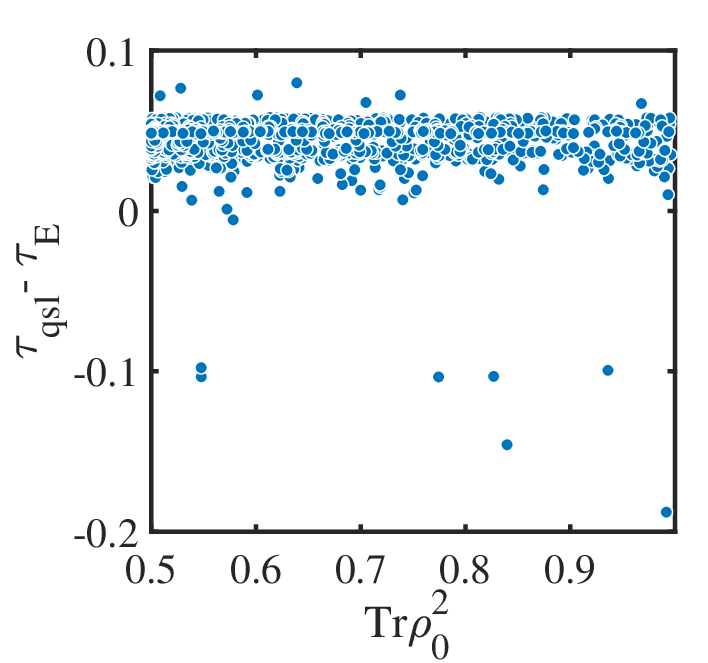}
\caption{The purity of initial state vs $\tau_{qsl}-\tau_E$. We randomly generated $4\times 4$ diagonal hermitian matrix as the total Hamiltonian $H$ for a bipartite qubit system. The initial state of the system is the product state $\rho_S\otimes \rho_E$, where $\rho_S$ and $\rho_E$ are the two-dimensional density matrices. And the states we concerned about is attained by tracing out the irrelevant parts by using the partial trace: $\rho_t=Tr_E(U_t\rho_S\otimes\rho_EU_t^\dagger)$, where $U_t=exp[-iHt]$.}
\label{random}
\end{figure}

\begin{figure}[htbp]
\includegraphics[width=6.0cm]{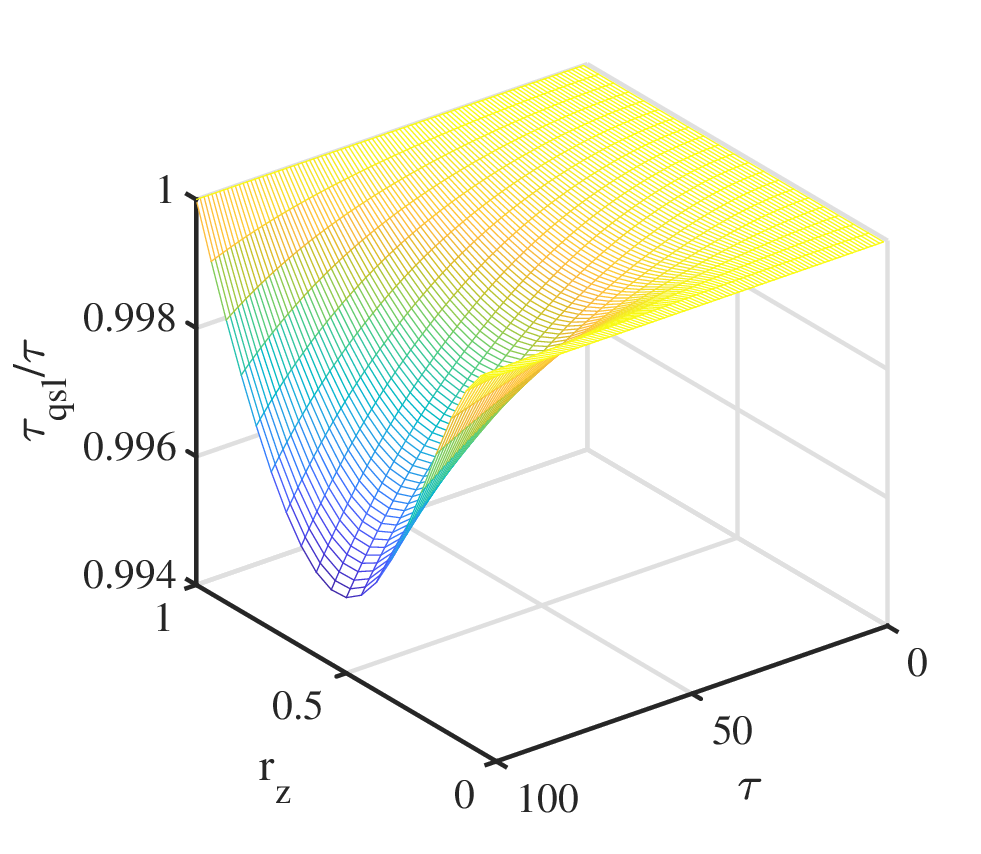}
\includegraphics[width=6.0cm]{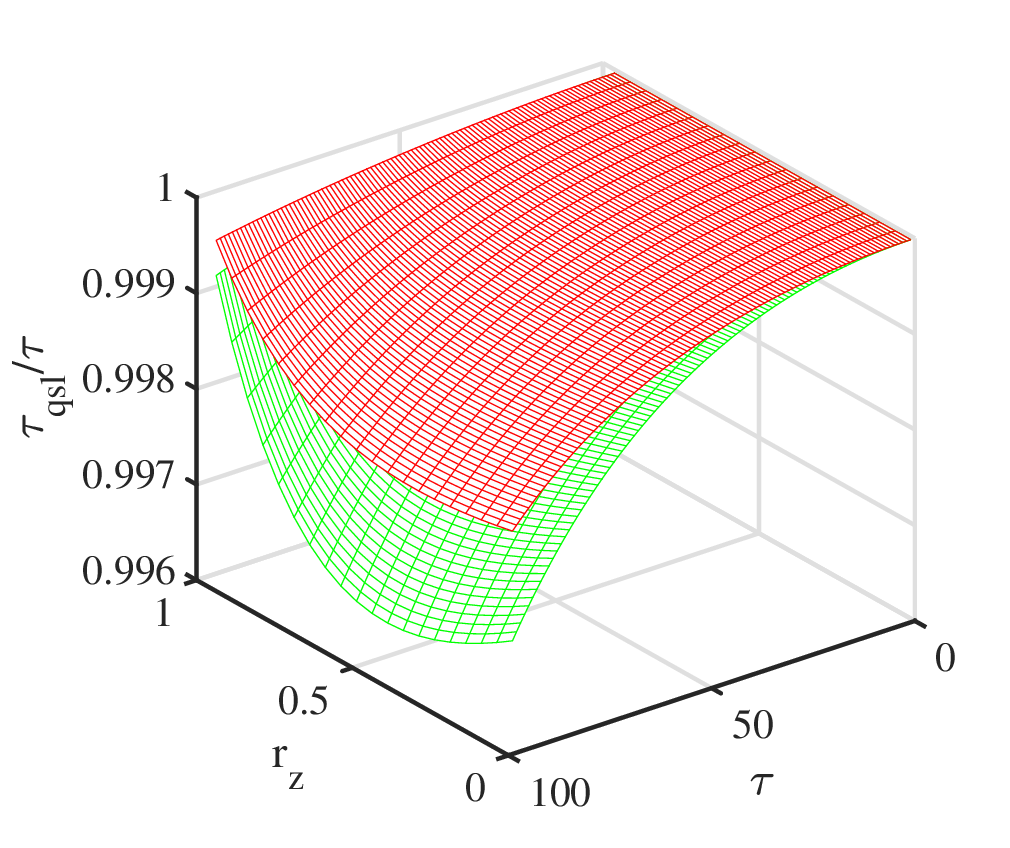}
\caption{(a) The ratio $\tau_{qsl}/\tau$ versus $\rho_{11}$ and $\tau$. Here $c=0.5$, $p(t)=\ln(1+t/100)$.
(b) The ratio $\tau_{qsl}/\tau$ of our QSL and
that in Ref. \cite{campaioli2019tight}. The temperature-determined parameter $c$ is set as zero. The red surface represents
 our QSLT and the green one corresponds to that in Ref. \cite{campaioli2019tight}.}
\label{equilibrium}
\end{figure}

However, to demonstrate the tightness in general nonunitary cases,  we sample $1000$ randomly generated  dynamics process for qubit systems to calculate  $\tau_{qsl}-\tau_E$ in Fig. \ref{random}.
One can find that most of the given examples show our bound is tight, but some demonstrate  $\tau_E$ is tighter than ours. 
For an analytical demonstration, let's consider a qubit
interacting with a large bath. The Kraus operators are given as \cite
{10}
\begin{eqnarray}
\begin{split}
K_{0}(t) =&\sqrt{c}(\sqrt{1-p(t)}\left\vert 0\right\rangle \left\langle
0\right\vert +\left\vert 1\right\rangle \left\langle 1\right\vert ),
\label{equ_master} \\
K_{1}(t) =&\sqrt{c}\sqrt{p(t)}\left\vert 1\right\rangle \left\langle
0\right\vert , \\
K_{2}(t) =&\sqrt{1-c}(\sqrt{1-p(t)}\left\vert 1\right\rangle \left\langle
1\right\vert +\left\vert 0\right\rangle \left\langle 0\right\vert ), \\
K_{3}(t) =&\sqrt{1-c}\sqrt{p(t)}\left\vert 0\right\rangle \left\langle
1\right\vert ,
\end{split}
\end{eqnarray}%
where $c$ is a parameter determined by the temperature of the bath and $%
p(t) $ is some increasing function of time $t$ describing the evolution
path. Suppose the initial state $\rho _{0}=(1-\rho _{11}(0))\left\vert
0\right\rangle \left\langle 0\right\vert +\rho _{10}(0)\left\vert
0\right\rangle \left\langle 1\right\vert +\rho _{10}^{\ast }(0)\left\vert
1\right\rangle \left\langle 0\right\vert +\rho _{11}(0)\left\vert
1\right\rangle \left\langle 1\right\vert $, the density operator can be
easily obtained as
\begin{equation}
\rho _{t}=\left(
\begin{matrix}
1+(\rho _{11}(0)-c)p(t)-\rho _{11}(0) & \sqrt{1-p(t)}\rho _{10}(0) \\
\sqrt{1-p(t)}\rho _{10}^{\ast }(0) & (c-\rho _{11}(0))p(t)+\rho _{11}(0)%
\end{matrix}%
\right).
\end{equation}%
For the convenience of computation, we adopt the Bloch representation to describe the initial state $\rho_{11}=(1-r_z)/2$ and $\rho_{10}$ is selected for ensuring the pure initial state. The ratio  $\tau _{qsl}/\tau $ of QSL time to the actual evolution time is plotted in Fig. \ref{equilibrium} (a), which indicates that the QSLT $\tau_{qsl}$ is less than the actual evolution time $\tau$ for most of the initial states, but is attainable if  $\rho_{11}=0$ and $\rho_{11}=c$. In fact, one can find that $\rho_{11}=0$ or $c$ is the exact condition that can ensure
 $\dot{\protect\rho}_t=%
\dot{\protect\beta}(t)C$, where $C$ is time-independent, which is also the equivalent condition of geodesics dynamics. Additionally, we compare our QSLT with that proposed in
 Ref. \cite{campaioli2019tight} by
the ratio $\tau _{qsl}/\tau $ in Fig. \ref{equilibrium} (b). It is obvious that our
QSLT is larger (tighter) than the QSLT in Ref. \cite{campaioli2019tight}  .

Since the combination approach has been widely used in establishing tighter QSLT \cite{deffner2013quantum}, combining the different QSLs could provide a tighter bound for the evolution time. Namely, a combination QSLT form  as $\tau_{qsl}^{comb}=\left\{\tau_{qsl},\tau_E\right\}$ should be a good QSLT for an open system.

\section{Discussion and conclusion}
We have established a quantum speed limit for the open system by an
intuitive geometrical picture. For any
initial state, one can always find corresponding dynamics to achieve the "fastest" evolution along the geodesic. We found the general condition
for dynamics to saturate our QSL. By evidence, we consider the evolutions of a quantum state undergoing the generalized amplitude damping channel and the dephasing channel, which verify the attainability of our QSL when the decay rates are monotonic. But for the dynamics with the non-monotonic decay rate, such as the case of non-Markovian dynamics, the bound is unsaturated. We compare our QSLT with the tight one $\tau_E$ presented in Ref. \cite{campaioli2019tight}.  We show our bound is tighter than $\tau_E$ for pure initial state governed by unitary (or close to unitary) evolution.  Besides, we sample $1000$ non-unitary dynamics for qubit systems,  and find that for most cases our bound is tighter than t$\tau_E$, but for some other cases the result is contrary, which implies the combination of the two QSLTs should be a tighter bound.   In summary,  we have presented attainable bound for QSLT,  which provide a different understanding of QSLT.
\section*{Acknowledgements}
This work was supported by the National Natural Science Foundation of China under Grants No.12175029, No.
12011530014 and No.11775040.

\section*{Appendix A}
In this section, we show a proof to verify that Eq. (\ref{geoeqn}) is the geodesics. Let $\Delta \rho _{\tau 0}=\rho _{\tau }-\rho _{0}$, then $%
\rho _{t}$ can be rewritten as $\rho _{t}=\rho _{0}+\beta (t)\Delta \rho
_{\tau 0}.$ The derivation of $\rho _{t}$ reads $\dot{\rho}_{t}=\dot{\beta}%
(t)\left( \rho _{\tau }-\rho _{0}\right) $, which is exactly the equation (%
\ref{geoeqn}). Solving the differential equation (\ref{geoeqn}), one will
obtain Eq. (\ref{geodesic}).

We will show that Eq. (\ref{geoeqn}) is the geodesic. One can easily
find
\begin{equation}
\begin{split}
Tr\dot{\rho}_{t}^{2} =& \dot{\beta}(t)^{2}Tr(\Delta \rho _{\tau })^{2}, \\
Tr\rho _{t}^{2} =& Tr\rho _{0}^{2}+\beta (t)^{2}Tr(\Delta \rho _{\tau
0})^{2}+\\&2\beta (t)Tr\rho _{0}\Delta \rho _{\tau 0}, \\
Tr\rho _{t}\dot{\rho}_{t} =&\dot{\beta}(t)[Tr\rho _{0}\Delta \rho _{\tau
0}+\beta (t)Tr(\Delta \rho _{\tau 0})^{2}],
\end{split}
\end{equation}%
so the average evolution speed can be calculated as

\begin{equation}
\begin{split}
\left\vert v_{t}\right\vert =& \frac{\sqrt{Tr\rho _{t}^{2}Tr\dot{\rho}%
_{t}^{2}-(Tr\rho _{t}\dot{\rho}_{t})^{2}}}{Tr\rho _{t}^{2}} \\
=& \frac{|\dot{\beta}(t)|\sqrt{Tr\rho _{0}^{2}Tr(\Delta \rho _{\tau
0})^{2}-(Tr\rho _{0}\Delta \rho _{\tau 0})^{2}}}{Tr\rho _{0}^{2}+\beta
^{2}(t)Tr(\Delta \rho _{\tau 0})^{2}+2\beta (t)Tr\rho _{0}\Delta \rho _{\tau
0}} \\
=& \frac{\dot{\beta}(t)\frac{Tr(\Delta \rho _{\tau 0})^{2}}{\sqrt{Tr(\Delta
\rho _{\tau 0})^{2}Tr\rho _{0}^{2}-(Tr\rho _{0}\Delta \rho _{\tau 0})^{2}}}}{%
1+[\frac{\beta (t)Tr(\Delta \rho _{\tau 0})^{2}+Tr\rho _{0}\Delta \rho
_{\tau 0}}{\sqrt{Tr(\Delta \rho _{\tau 0})^{2}Tr\rho _{0}^{2}-(Tr\rho
_{0}\Delta \rho _{\tau 0})^{2}}}]^{2}} \\
=& \frac{d}{dt}\arctan \frac{\beta (t)Tr(\Delta \rho _{\tau 0})^{2}+Tr\rho
_{0}\Delta \rho _{\tau 0}}{\sqrt{Tr(\Delta \rho _{\tau 0})^{2}Tr\rho
_{0}^{2}-(Tr\rho _{0}\Delta \rho _{\tau 0})^{2}}},
\end{split}%
\end{equation}%
where $\dot{\beta}(t)>0$ is considered since $\beta (t)$ is a monotonic
function with $\beta (0)=0$ and $\beta (\tau )=1$. The evolution path is

\begin{equation}
\begin{split}
\int_{0}^{\tau }\left\vert v_{t}\right\vert dt& =\arctan \frac{Tr(\Delta
\rho _{\tau 0})^{2}+Tr\rho _{0}\Delta \rho _{\tau 0}}{\sqrt{Tr(\Delta \rho
_{\tau 0})^{2}Tr\rho _{0}^{2}-(Tr\rho _{0}\Delta \rho _{\tau 0})^{2}}} \\
& -\arctan \frac{Tr\rho _{0}\Delta \rho _{\tau 0}}{\sqrt{Tr(\Delta \rho
_{\tau 0})^{2}Tr\rho _{0}^{2}-(Tr\rho _{0}\Delta \rho _{\tau 0})^{2}}}.
\end{split}%
\end{equation}%

A simple calculation can show that $\cos \int_{0}^{\tau }\left\vert
v_{t}\right\vert dt=Tr\rho _{0}\rho _{\tau }/(\sqrt{Tr\rho _{0}^{2}}\sqrt{
Tr\rho _{\tau }^{2}})$, which means Eq. (\ref{geodesic}) is the geodesic.
The proof is finished. \hfill {} $\square $

\section*{Appendix B}
Here we provide an example to illustrate that the system Hamiltonian can drive the evolution trajectory to deviate from the geodesics. Consider the master equation (\ref{mas1}) in the Schr\"{o}dinger picture as
\begin{equation}\label{schrodinger}
\begin{split}
  \dot{\rho}_t=&-i[H_\theta,\rho_t]+\frac{\gamma}{2}\left(\Sigma_-\rho_t\Sigma_+-\Sigma_+\Sigma_-\rho_t-\rho_t\Sigma_+\Sigma_-\right),\\
    H_\theta=&\frac{\Omega_L}{2}\left(\cos \theta \sigma_z+\sin \theta \sigma_x\right),
\end{split}
\end{equation}
where $\gamma$ is the time-independent decay rate, $\Sigma_{\pm}=U\sigma_{\pm}U^\dagger$ with $U=\cos\frac{\theta}{2}I-i\sin\frac{\theta}{2}\sigma_y$, $H_\theta$ is the parameter $\theta$ determined Hamiltonian with the eigenfrequency of $\Omega_L=\sqrt{\epsilon^2+\Omega^2}$, and $\epsilon$ and $\Omega$ denote the energy level difference and tunneling coupling, respectively. The solution of Eq. (\ref{schrodinger}) is presented in Ref. \cite{Lan_2022} as
\begin{equation}\label{solu_lan}
  \rho_t=\frac{1}{2}\left(
  \begin{matrix}
    1+r_z(t) & r_x(t)-ir_y(t) \\
    r_x(t)+ir_y(t) & 1-r_z(t)
  \end{matrix}\right)
\end{equation}
with
\begin{eqnarray}\label{rx}
  &r_x(t)=e^{-\frac{\gamma}{2}t}\bigg\{\left[ \sin^2\theta e^{-\frac{\gamma}{2}t}+\cos^2\theta \cos(\Omega_L t)\right]r_x(0)\notag\\
&  -\cos\theta\sin(\Omega_Lt)r_y(0)
  +\sin\theta\cos\theta\left[e^{-\frac{\gamma}{2}t}\right.\notag\\
  &-\left.\cos(\Omega_Lt)\right] r_z(0)\bigg\}
  +\sin\theta\left[e^{-\gamma t}-1\right]
\end{eqnarray}
\begin{equation}\label{ry}
\begin{split}
r_y(t)=e^{-\frac{\gamma}{2}t}\big[\cos\theta\sin(\Omega_Lt)r_x(0)+\cos(\Omega_Lt)r_y(0)\\
-\sin\theta\sin(\Omega_Lt)r_z(0)\big]
\end{split}
\end{equation}

\begin{eqnarray}\label{rz}
  &r_z(t)=e^{-\frac{\gamma}{2}t}\bigg\{\sin\theta\cos\theta\left[e^{-\frac{\gamma}{2}t}-\cos(\Omega_Lt)\right]r_x(0)\notag\\
 & +\sin\theta\sin(\Omega_Lt)e^{-\frac{\gamma}{2}t}r_y(0)
  +\left[ \cos^2\theta e^{-\frac{\gamma}{2}t}\right.\notag\\
  &+\left.\sin^2\theta \cos(\Omega_L t)\right]r_z(0)\bigg\}
  +\cos\theta\left[e^{-\gamma t}-1\right]
\end{eqnarray}

According to Eq. (\ref{geodesic_1}), the geodesics dynamics can be expressed as a product of real time-dependent factor and a time-independent hermitian matrix with zero-trace, hence the time-dependent imaginary factor of the non-diagonal entries should be vanish, it means that $r_y=0$, i.e., $r_y(0)=0$, $r_x(0)=r_z(0)$, $\theta=\pi/4$, one can immediately obtain that Eq. (\ref{solu_lan}) is the geodesics due to $r_x(t)=r_y(t)$, and $[H_\theta,\rho_t]=0$. That is, for this model, the system Hamiltonian drives the evolution trajectory deviate the geodesics except for the case when $\rho_t$ is commutative with the Hamiltonian. In the general cases, the presence of the time-dependent imaginary factor of the non-diagonal entries of the dynamics matrix always lead to a non-geodesics evolution.

\section*{References}
\bibliography{ref2}
\end{document}